\documentclass[aps,prb,twocolumn,showpacs,amssymb,superscriptaddress]{revtex4-1} 
\usepackage{graphicx}
\usepackage{graphicx}
\usepackage{bm}
\usepackage{amsmath}
\usepackage{amssymb}

\newcommand{\ket}[1]{\ensuremath{\left|{#1}\right\rangle}}
\newcommand{\bra}[1]{\ensuremath{\left\langle{#1}\right|}}
\newcommand{\braket}[1]{\ensuremath{\left\langle{#1}\right\rangle}}

\begin{document}

\title{Detecting ground state qubit self-excitations in circuit QED:\\ slow quantum anti-Zeno effect}

\author{C. Sab\'in}
\email{csl@iff.csic.es}
\affiliation{Instituto de F\'{\i}sica Fundamental, CSIC,
  Serrano 113-B, 28006 Madrid, Spain}

\author{J. Le\'on}
\affiliation{Instituto de F\'{\i}sica Fundamental, CSIC,
  Serrano 113-B, 28006 Madrid, Spain}

\author{J.~J. Garc{\'\i}a-Ripoll}
\affiliation{Instituto de F\'{\i}sica Fundamental, CSIC,
  Serrano 113-B, 28006 Madrid, Spain}

\begin{abstract}
In this  work we study an ultrastrong coupled qubit-cavity system subjected to slow repeated measurements. We demonstrate that even under a few imperfect measurements it is possible to detect transitions of the qubit from its free ground state to the excited state. The excitation probability grows exponentially fast in analogy with the quantum anti-Zeno effect. The dynamics and physics described in this paper is accessible to current superconducting circuit technology.
\end{abstract}


\maketitle
\section{Introduction}

In circuit QED the model of a two-level system interacting with one or more harmonic oscillators can be implemented combining a superconducting qubit with a microwave resonator or a transmission line~\cite{blais04,wallraff04,chiorescu04}. Compared to experiments in Quantum Optics with microwave cavities~\cite{raimond01,walther06} or with trapped ions~\cite{leibfried03}, the superconducting circuit experiments have one important advantage: the strength of the qubit-photon coupling. The fact that superconducting resonators and superconducting qubits follow essentially the same physical laws makes it possible not only to reach the strong coupling regime~\cite{wallraff04,chiorescu04}, in which multiple Rabi oscillations are possible within the decoherence of the cavity or the qubit, $g \gg \kappa, \gamma,$ but also entering the ultrastrong coupling regime, $g \sim \omega,$ in which the internal and interaction energies become similar~\cite{niemczyk10,forn-diaz10}. In this new regime the dynamics is very fast and the usual approximations such as the Rotating Wave Approximation (RWA) in the Jaynes-Cummings model break down~\cite{irish05,irish07,ashhab10}.

One of the most astounding predictions of the ultrastrong coupling regime is that a single qubit can distort its electromagnetic environment, giving rise to a ground state in which the qubit is dressed with photons. As we will show in the following sections, in the case of a qubit and a single harmonic oscillator, this translates into a state which is a superposition of a deexcited qubit and a vacuum, with other states in which the qubit, the oscillator or both are populated with excitations and photons, respectively~\cite{irish07,casanova10,hausinger10}. This is a completely non-RWA effect which requires large values of the coupling to be observed. More precisely, the excitation probability grows approximately as $p_e \propto (g/\omega)^2$ and $g$ has to become comparable to the energies of a photon, $\hbar \omega,$ or of a qubit, $\hbar\omega_0,$ making the interaction dynamics both very strong and very fast. From the experimental point of view it would thus seem unfeasible to probe a physics that takes place at speeds of $\omega\sim1-10$ GHz, while the typical measurement apparatus in circuit-QED have response times which are much slower, of about 50 ns. There are four routes to escape this problem: making the ultrastrong coupling switchable by design~\cite{peropadre10}, dynamically turning it off by external drivings~\cite{hausinger10b}, engineering faster measurement apparatus or looking for new ways to extract information out of slow measurement devices.

In this work we take the slow route,  showing that is possible to extract valuable information from the fast dynamics of the system with current measurement technologies. We will study what happens to an ultrastrongly coupled qubit-cavity system when the qubit is subject to repeated measurements by a detector with a slow repetition rate that is only capable of performing weak measurements of the state of the qubit. The main goal is to detect the qubit in its excited state starting from the ground state of the system. The first measurement has already a small probability of success, as commented in the previous paragraph. In case of failure the system is projected to a non-equilibrium state which rapidly exhibits a dynamics with an oscillatory probability of excitation, mainly due to non-RWA transitions from the ground state of the qubit $\ket{g}$ to the excited one $\ket{e}$. By means of performing repeated measurements, we will show that the detector is able to probe these usually considered as ``virtual''  excitations of the qubit and the cavity and at the same time reveal information of the interaction model. More precisely, the repeated measurements accumulate information exponentially fast and behave like an anti-Zeno effect \cite{saveriozeno} in which the qubit is projected onto its excited state, revealing those ground-state excitations that we were looking for. We show that  this anti-Zeno ``decay'' $\ket g \rightarrow \ket e,$ is very efficient and does only require a \textit{short number of repeated measurements} with a repetition rate which is much slower than in the standard anti-Zeno effect.

Like previous proposals for probing the ultrastrong coupling limit~\cite{lizuain10, cqedsabin}, the anti-Zeno dynamics in this work is supported by the counter-rotating terms in the qubit-resonator interaction, using as seed the ground state excitations of these systems. The phenomenon is absent in the limit of RWA in Jaynes-Cummings models. Let us remark that the non-RWA effects are being extensively studied not only in the ultrastrong coupling regime of circuit-QED but also in other fields like Quantum Optics \cite{almutbeige}. Models of repeated measurements on superconducting qubits were considered for instance in Ref. \cite{calarcoonofrio} and  have been implemented in the lab \cite{measurementsqubits1, measurementqubits2}.

The structure of the text is as follows. In Sect. II we will show that the eigenstates of the hamiltonian, and in particular the ground state of a qubit-cavity system in the ultrastrong coupling regime are not separable, $\ket{g,2n}$ or $\ket{e,2n+1},$  but linear combinations of these vacua and excitations. More precisely, the qubit-resonator ground state contains a contribution of $\ket{e,1}$ which grows with the coupling strength and becomes relevant in the ultrastrong coupling regime, $g\sim \omega.$ We will see that after a few ideal periodic projective measurements of the qubit state, the probability of finding that it is in the state $\ket g$ tends quickly to 0, even if an uncertainty in the time taken by the measurement is considered. In section II. E we will consider a realistic model of  measurement  in which large  amounts of errors are allowed, showing the robustness of our method. Section II.F is devoted to the analysis of the role of relaxation and dephasing. We conclude in section IV with a summary of our results.

\section{Detecting self-excitations of the vacuum}

\subsection{The Dicke model}

We will consider the following Hamiltonian, corresponding to a qubit-cavity system
\begin{equation}
\label{eq:hamiltonian}
H = H_0 + g H_I =
\hbar \omega a^\dag a+ \frac{\hbar\omega_0}{2}\sigma^z + \hbar g \sigma^x (a + a^\dagger),
\end{equation}
where $\hbar \omega_0$ is the energy splitting between the two levels of the qubit $\ket e$ and $\ket g,$ $\omega$ the frequency of the photons in the cavity or resonator field and $g$ the coupling strength.

In the weak and strong coupling regimes, in which the coupling $g \ll \omega,\omega_0$ is only compared to the decay rates of the cavity and the qubit, one may treat $H_I$ as a small perturbation on top of the bare qubit and resonator states. In this limit the counter-rotating terms $a^\dagger\sigma^+, a\sigma^-$ average out, and the total Hamiltonian becomes equivalent to the Jaynes-Cummings model, whose ground state is a separable combination of the qubit ground state and a cavity vacuum, $\ket{g,0}.$

\begin{figure}[t]
  \centering
  \includegraphics[width=0.8\linewidth]{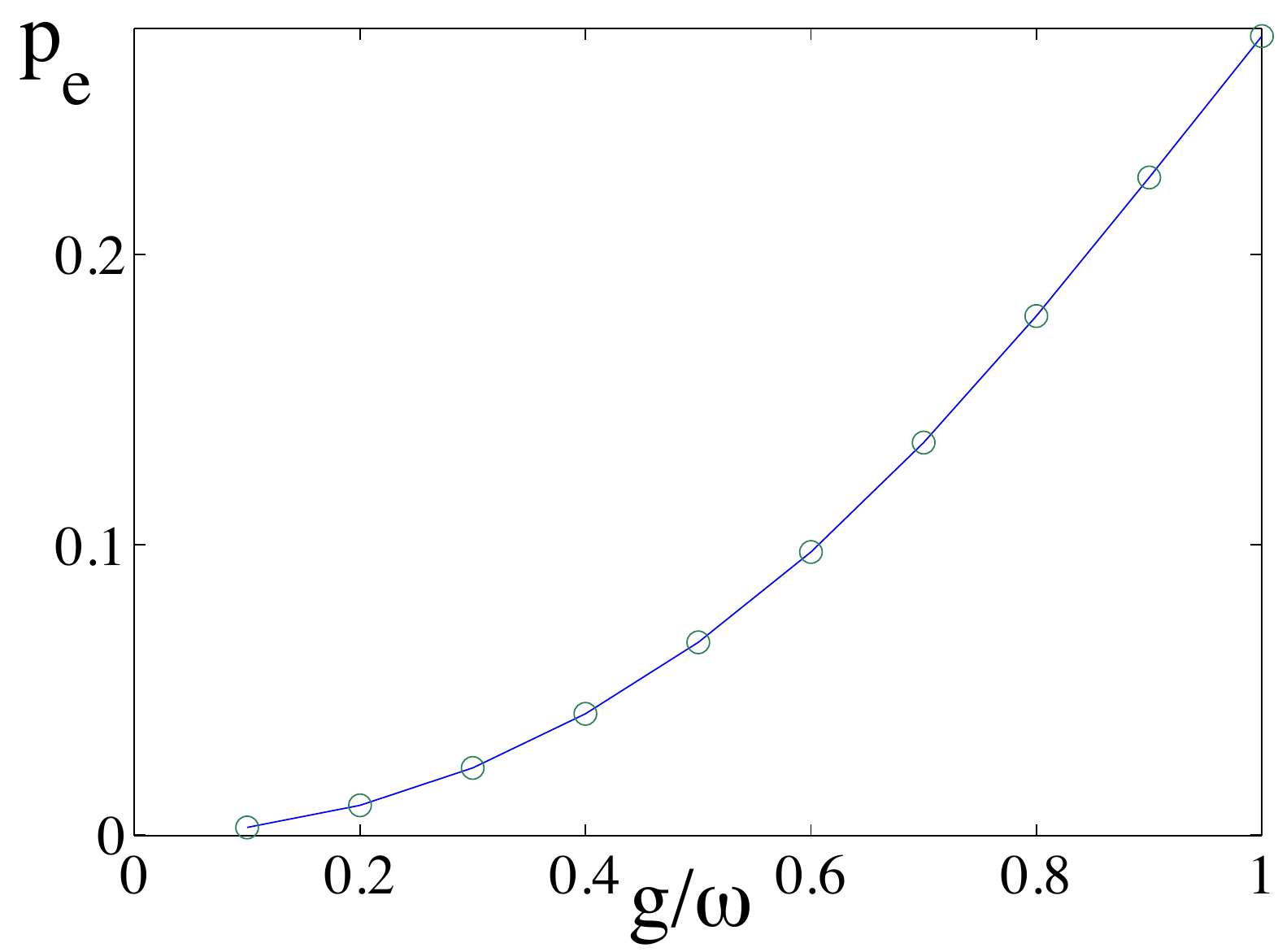}
  \caption{(Color online) Probability of excitation for of a qubit $p_e$ (blue,solid) vs. the dimensionless ratio $g/\omega,$ for a qubit-resonator system [Eq.~(\ref{eq:hamiltonian})] in the ground state of a ultrastrong coupled limit $\omega=\omega_0=1$ GHz. This line is undistinguishable from a quadratic fit (green,dashed). }
  \label{fig:groundstate}
\end{figure}

In this work we are interested  however in the ultrastrong coupling regime, in which $g$ approaches the qubit and photon frequencies, $\omega$ and $\omega_0.$ In this case it is more convenient to look at the state space in the language of parity subspaces~\cite{casanova10}, and treat $H_0$ and $H_I$ on equal footing. Within this picture, the Hilbert space splits up in two different chains of states coupled by $H_I,$ and in particular the ground state of the system becomes a linear combination of states in the even parity sector
\begin{equation}
\label{eq:groundstate}
\ket{G}= c_0\ket{g0}+c_1\ket{e1}+c_2\ket{g2}+c_3\ket{e3}+ \ldots
\end{equation}
where the coefficients $c_i$ depend on $g,\omega,\omega_0.$

\subsection{Detecting excitations with one measurement}

One of the goals of this paper is design a protocol for measuring the tiny excitations in the ground states ---$|c_1|^2+|c^3|^2+|c_5|^2+\ldots$ in Eq.~(\ref{eq:groundstate})---. Let us assume for now that we have a good measurement apparatus and that we perform a single measurement of the qubit in the ground $\ket{G}$ of the system. In Fig.~\ref{fig:groundstate} we plot the probability of finding the qubit excited after just one measurement
\begin{equation}
p_{e}=\braket{\hat{P}_e}_G =: \braket{\ket{e}\bra{e}}_G \label{eq:pe}
\end{equation}
against different values of the coupling strength, assuming always $\omega=\omega_0$ and $g/\omega \le1$. For the strongest couplings the values of $p_e$ are sizable. Moreover, we have:
\begin{equation}
p_{e}=\lambda \frac{g^2}{\omega^2}, \,\, (\omega=\omega_0,\,\frac{g}{\omega}\le1) \label{eq:pequadratic}
\end{equation}
This quadratic behavior comes as no surprise. The main contribution to $p_{e}$ is $|c_1|^2.$ If we think of $\ket G$ as the free vacuum $\ket{g,0}$ dressed by the interaction $H_I,$ then $|c_1|$ may be computed from perturbation theory in interaction picture, the leading term  being proportional to $\left|\braket{e,1|H_I|g,0}\right|^2.$ It is interesting to see how these contributions quickly grow as $g$ approaches $\omega,$ but that at the same time the signal in current experiments with $10\%$ coupling strengths, might have a too small excitation signal to be accurately detected.

This work is born from the idea that perfect projective measurements in c-QED might be too difficult, as existing measurement apparatus may be too slow or not have enough sensitivity to capture those excitations. The constraint of time is found, for instance, in flux qubit measurement devices based on SQUIDs, which roughly work as follows: A very short current pulse is sent to the SQUID, instantaneously changing its potential from a periodic function to a washboard potential. In this brief period of time, one of the flux qubit states which is sitting inside the SQUID may provide, through its intrinsic current and flux, enough additional energy for the SQUID to tunnel into a voltage state. This stochastic process is random in time and does no have a 100$\%$ success rate. Moreover, it requires an additional sustained current that keeps the SQUID in that voltage state during an integration time large enough for the electronics to realize that the measurement succeeded. Adding the excitation and integration phases, the best experimental setups bring the detection time down to tens of nanoseconds, which is still slower than the qubit-resonator dynamics --$1.6$ ns for a $600$ MHz coupling, and much shorter for the qubit and resonator periods, $1/\omega.$

An additional complication of the ultrastrong coupling limit is that an arbitrary measurement device might not have enough good coupling to either the qubit or the resonator in the ultrastrong coupling regime. If we assume that both quantum systems interact so strongly that their eigenstates are highly entangled states with large energy gaps, $\sim g,\omega,\omega_0,$ the detector could have problems coupling to those states and breaking their energy level structure. In other words, the measurement device couples through an operator, $\sigma^z,$ which typically represents a perturbation of the qubit-resonator model, and if that perturbation, which aims at breaking the linear combinations~(\ref{eq:groundstate}), is not strong enough, it might not extract any information from the system, or the amount of information might be reduced, becoming an off-resonant, weak dispersive measurement.

\begin{figure}[t!]
  \centering
  \includegraphics[width=0.9\linewidth]{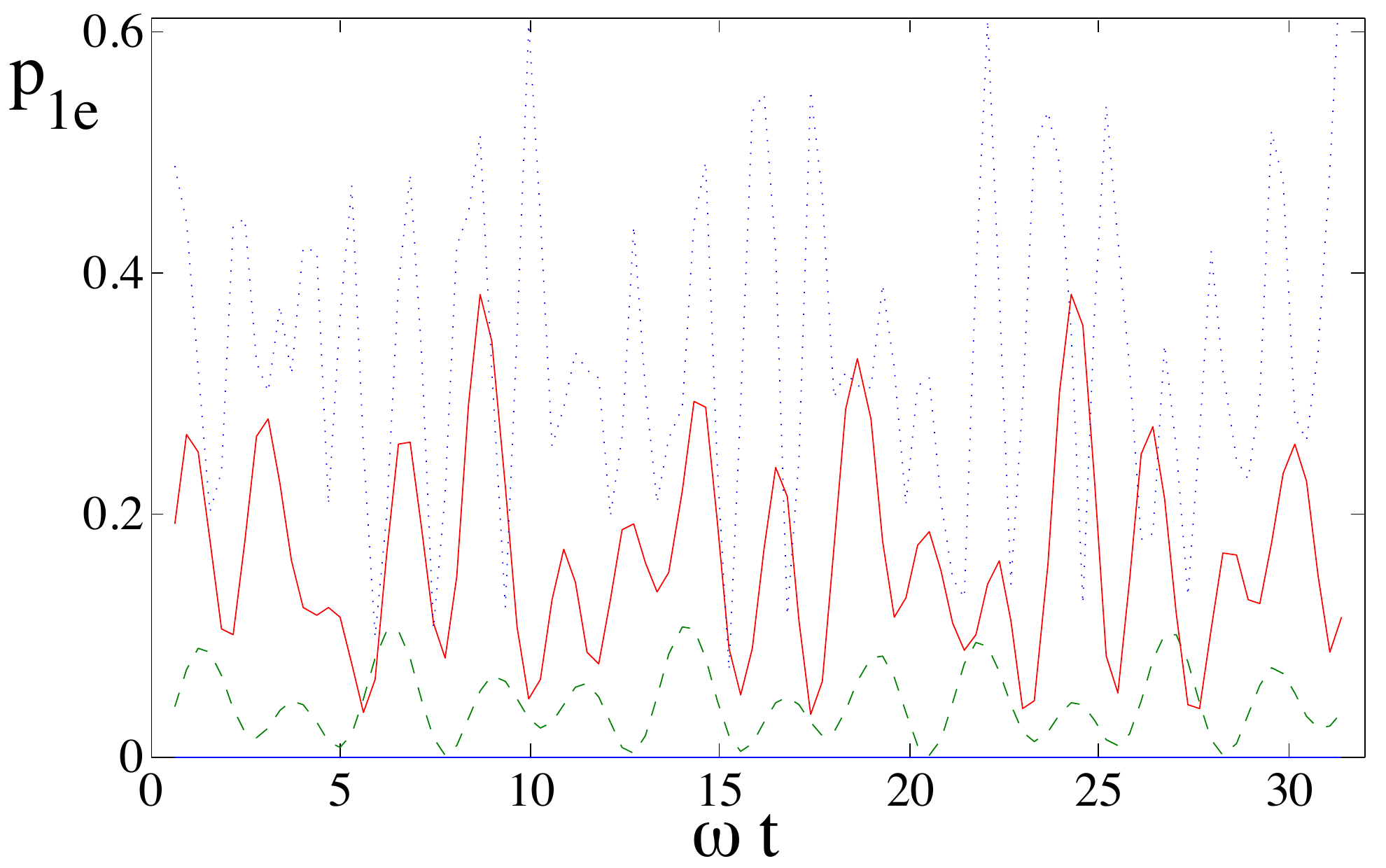}
  \caption{(Color online) After the qubit has been measured once, the qubit in the qubit-resonator system is left in a non-equilibrium state, $\ket{\Psi}.$ Here we plot the probability of excitation for the qubit $p_{1e}$ as a function of the dimensionless time $\omega t,$ shortly after that measurement. We show three situations, $\omega=\omega_0= 1$ GHz and $g/\omega=1/3 $ (dashed), $g/\omega=2/3$ (solid) and $g/\omega=1$ (dotted), which exhibit fast dynamics.} \label{fig:onemeasurement}
\end{figure}

All these considerations brought us to the idea of using more than one measurement steps in the same experiment, with the aim of increasing the amount of information that it is extracted from the same state. This can be done because the kind of measurements done in experiments are non-destructive: the same qubit can be continued to be measured at another time. It is true, however, that the interval between measurements might carry a strong, fast and almost chaotic dynamics~[Fig.~\ref{fig:onemeasurement}], arising from the fact that the measurement brings the system into a non-equilibrium state, even if it did not produce any information. We will show that this is not a limitation, but a plus, and that the repeated measurements may characterize the intermediate dynamics.

\subsection{Repeated measurements: survival probability}
\label{sec:repeated}

If we measure the qubit once, the measurement apparatus does not click and we are working with a perfect projective measurement, we conclude that the qubit-resonator system has been projected onto the state
\begin{equation}\label{eq:projection}
  \ket{\Psi} = \sum_{n}\hat{c}_{2n} \ket{g,2n},
\end{equation}
which is a (normalized) linear combination of deexcited qubits and some photons in the cavity. By measuring the ground state in an improper basis, we have created a non-stationary state that will evolve very quickly, with frequencies that are close to $g,\omega$ and $\omega_0.$ Lacking any other relaxation mechanism than the cavity and qubit decoherence times, these oscillations will be sustained for a large period of time, causing the qubit to get reexcited multiple times. The excitation probability
\begin{equation}
  p_{1e}(t,0)=\braket{\Psi(t)\left|\hat{P}_e\right|\Psi(t)}, \label{eq:p1e}
\end{equation}
may be computed from the initially measured state as
\begin{equation}
  \ket{\Psi(t)} \propto e^{-iHt} (1-\hat{P}_e) \ket{G}.
\end{equation}
As Fig.~\ref{fig:onemeasurement} shows, $p_{1e}$  exhibits very fast oscillations, but  also  average to a nonzero value, which is always close to the ground state excitation probability of the qubit, $p_e = \sum_n |c_{2n+1}|^2.$ Consequently, if we perform a second measurement at a later time $t_1$ we will have again a certain probability of success $p_{1e}(t_1,0)$ of detecting the state $\ket{e},$ and a certain probability of failure $p_{1g}(t_1,0)=1-p_{1e}(t_1,0).$ In the latter case the system is projected to a new state with a new time dependent probability $p_{2e}(t_2,t_1),$ and so on. After a few measurements we can define the survival probability as the probability that we have never detected a state $\ket{e}$ in the qubit
\begin{equation}
P_g^N=p_g p_{1g} (t_1,0) p_{2g}(t_2,t_1)...p_{Ng}(t_N,t_{N-1}).\label{eq:survival}
\end{equation}
A key idea in the interpretation of this formula is the fact that the intermediate probabilities $p_{ng}$ are on average very similar, and almost independent of the timespan among measurements. For the range of couplings that are within intermediate reach in experiments, $g/\omega \sim 0.1 - 1,$ we have verified numerically and perturbatively that this probability is well approximated by a quadratic law
\begin{equation}
p_{ng} \sim 1- \chi_n \frac{g^2}{\omega^2} 
\end{equation}
with minor differences among realizations, $\chi_n.$  The accumulation of products in Eq.~(\ref{eq:survival}) leads to an approximately exponential decrease of the survival probability
\begin{equation}
  P_g^N \sim \prod_{n=1}^N \left(1- \chi_n \frac{g^2}{\omega^2}\right) \sim \exp \left(-N\bar\chi\frac{g^2}{\omega^2}\right),
  \label{eq:exponential}
\end{equation}
as long as $\bar\chi\frac{g^2}{\omega^2}<<1$. This exponential behavior is typical of the so called anti-Zeno effect, in which repeated measurements of a quantum system accelerate the transition of a quantum system between two states. In our case the repeated measurements are rather creating a non-unitary evolution that excites the qubit from $\ket{g}$ to $\ket{e}$ using as seed the nonzero excitation probability $p_{1e}=\sum_n |c_{2n+1}|^2$ which is present in the equilibrium state of the qubit-resonator system. This last point is particularly important because this anti-Zeno evolution is impossible when the ground state of the qubit and the resonator is the vacuum $\ket{g,0}.$ In this case $g/\omega$ is so small, and $p_{1e}$ so close to zero, that all measurements will give no signal at all and the qubit will remain in the state $\ket{g}$ for the duration of the experiment. As we will see in the following, there is a key difference between the effects described in this section and the standard anti-Zeno effect: we need only a few measurements and they can be widely spaced in time.

In the following sections we will summarize extensive numerical studies of the anti-Zeno dynamics. We have contrasted these with various semi-analytical methods, one of which, the use of truncated Hilbert spaces, helps us in understanding the reason for this behavior. For the range of couplings of current interest, $g/\omega \sim 0.1 - 1,$ it suffices to take two photons, and the ground, $\ket{G},$ plus the two excited states $\ket{E'}$, $\ket{E''}$ within the same parity subspace.  All states can be expanded as in Eq.~(\ref{eq:groundstate}) with coefficients $c_i,\,c_i',\, c_i''.$, as linear combinations of  $\ket{g0}$, $\ket{e1}$, $\ket{g2}$. After the first measurement, the qubit will end up in an excited state with probability $|c_1|^2$ and it will remain in the unexcited state with $|c_0|^2\simeq 1 - |c_1|^2,$ ending up in a combination
\begin{equation}
 \hat{P}_g\ket{G} = c_0 \ket{G} + c_0'\ket{E'} + c_0''\ket{E''}+\ldots
\end{equation}
The crudest approximation would be to neglect all excited state contributions and assume that after each measurement, either the state $\ket{e}$ is detected, or the system ends up in $\ket{G}.$ In this case the survival probability would be exactly exponential
\begin{eqnarray}
  P_g^N = (1 - |c_1|^{2} )\prod_{i=i}^N|c_0|^{2} = |c_0|^{2N+2}.
  \label{eq:truncated-exp}
\end{eqnarray}
In practice, however, the combined system does not end up only on the ground state, but gets excited state contributions from $\ket{E'},\ket{E''}.$ When we average the contributions over the period in which the measurement takes place, we find that already after the first measurement step, the excited states add up to the total probability, $\braket{p_{1e}}_T=|c_1|^2\,|c_0|^2+|c_1'|^2\,|c_0'|^2+|c_1^{''}|^2\,|c_0^{''}|^2+\ldots,$ enhancing the original behavior.

\begin{figure}[t]
  \centering
  \includegraphics[width=0.95\linewidth]{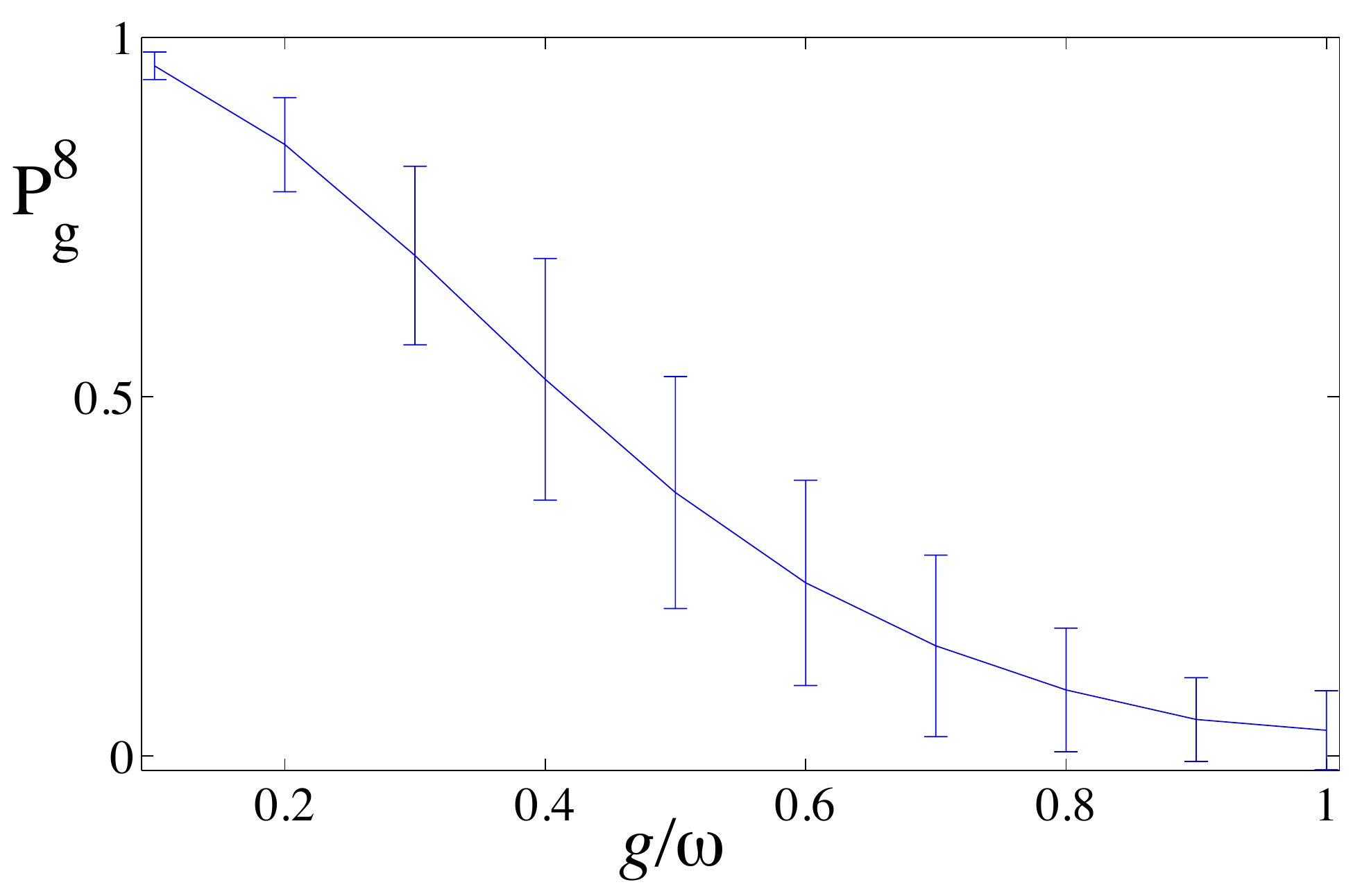}
  \caption{Survival probability after eight measurements $P_g^8$ vs. $g/\omega,$ with $\omega=\omega_0.$ The measurements are performed with periods $T_1,$ $T_2=\sqrt{2} T_1$ and averaged over 100 values of $T_1$ within the interval $2\pi [0.1,5].$ Note how the law approximates the Gaussian behavior in Eq.~(\ref{eq:exponential}).} \label{fig:8M}
\end{figure}

\subsection{Numerical experiments}

We have verified the anti-Zeno dynamics and the exponential law~(\ref{eq:exponential}) by means of exact numerical simulations in which we compute the outcome of repeated measurements on a qubit-resonator Dicke model~(\ref{eq:hamiltonian}). We will now explain the main results of this study.

From an experimental point of view it might be interesting to maximize the exponent $\bar\chi,$ optimizing the measurement repetition rate to hit all the maxima in the evolution of the excitation probability~[See Fig.~\ref{fig:onemeasurement}]. However we found that this is very difficult and demands a lot of precision on the measurement apparatus; for small errors or some measurement randomization this procedure drives the apparatus into exactly the opposite regime: always hitting the minima of excitation. Seeking a more robust, less demanding approach we opted for using two incommensurate periods, $T_1$ and $T_2 \simeq \sqrt{2}T_1,$ simulating measurement at times
\begin{equation}
  \label{eq:times}
  t_n \in \{T_1, T_1+T_2, 2T_1+T_2, 2T_1+2T_2,\ldots\},
\end{equation}
and at most optimizing the value of $T_1.$

\begin{figure}[t]
  \centering
  \includegraphics[width=\linewidth]{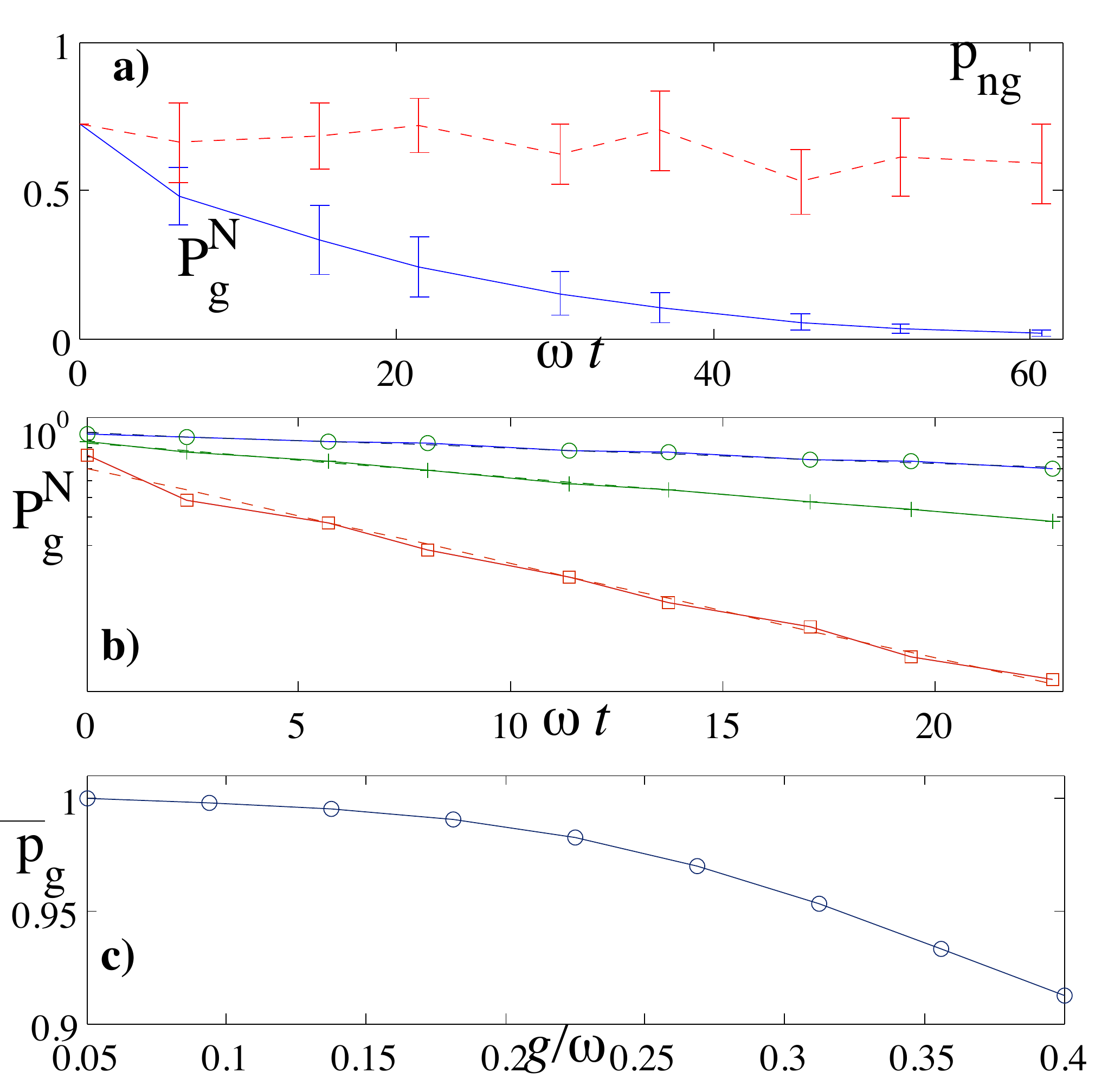}
  \caption{(a) Probability after the $n-$th single measurement, $p_{ng},$ (dashed) and accumulated survival probability, $P_g^N = \prod_{n=1}^N p_{ng}$ (solid) vs. dimensionless time $\omega t.$ We use $g/\omega=1$ and perform measurements with approximate periods $\omega T_1=\omega T_2/\sqrt{2}=2\pi,$ averaging over random perturbations of the actual measurement time, $t_n,$ within the interval $\omega t_n+[-0.2\pi,+0.2\pi].$ (b) Survival probability $P_g^N$ (solid lines) and the corresponding exponential fits (dashed lines) for $g/\omega=1/3$ (circles), $2/3$ (crosses) and $1$ (squares). (c) Mean value $\bar{p_g}$ (solid)  and the corresponding quadratic fit (dashed) vs. dimensionless coupling strength $\frac{g}{\omega}.$ All plots assume $\omega=\omega_0$ and (b,c) use $\omega T_1=3\pi/4$}
  \label{fig:8Mtime} 
\end{figure}

With this  approach, and exploring different values of $T_1,$ we have studied the survival probability and concluded that the exponential laws are really accurate. As shown in Fig.~\ref{fig:8M}, if we fix the total number of measurements to be $N=8$ and sample various periods, $T_1,$ we recover on average the Gaussian behavior $\exp(-N\bar\chi g^2/\omega^2)$ deduced in Eq.~(\ref{eq:exponential}). Instead of fixing the number of measurements, we can also study the same law and verify the exponential decay with respect to $N.$ This is shown in Figs.~\ref{fig:8Mtime}a-b, where we plot the accumulated survival probability, $P^N_g,$ as a function of time, and fit it against the same exponential~(\ref{eq:exponential}).

It is important to remark that the exponential decay is a robust signature that survives even when the measurement does not take place at  precise times, $t_n,$ from the list given before~(\ref{eq:times}). This has been verified by simulating multiple runs in which $t_n$ is randomly perturbed around its average value, and computing the survival probability. We want to remind the reader the importance of this robustness, because some measurement apparatus such as SQUIDs behave stochastically and produce a signal at a random time that can not be determined a-priori. The fact that the measurement protocol and the resulting physical behavior are independent of a precise control is encouraging.

The accuracy of the exponential law~(\ref{eq:exponential}) suggests that the survival probability of a single measurement remains constant throughout a single experiment, $p_g \simeq p_{ng}\;\forall n.$ This is qualitatively confirmed by Fig.~\ref{fig:8Mtime}a, where we show that these values oscillate around a mean one that is close to the average population of $\ket{g}$ in the ground state, i.~e. to $p_{g}.$ This suggests us to consider average values and approximate
\begin{equation}
  \bar{p}_{g}=\sum_n \frac{p_{ng}}{N}\simeq 1-\bar\chi\frac{g^2}{\omega^2}\label{eq:pgmean} 
\end{equation}
which has the expected quadratic behavior. This estimate is confirmed by Fig.~\ref{fig:8Mtime}c, where the quadratic fit is almost undistinguishable from the actual behavior.
\begin{figure}[t]
  \centering
  \includegraphics[width=0.95\linewidth]{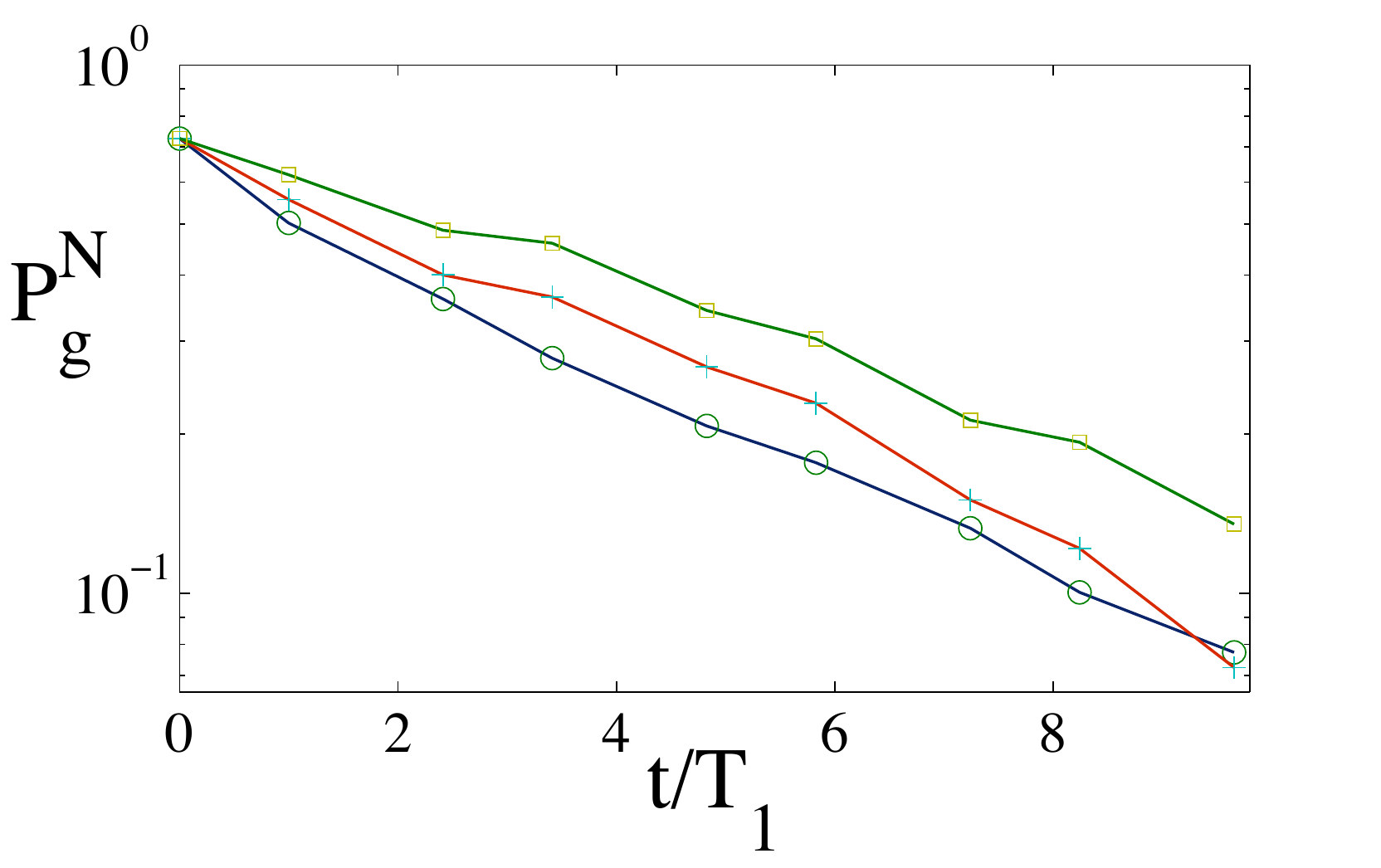}
  \caption{Survival probability $P_g^N$ vs. $ t/T_1$, with $g=\omega=\omega_0$ GHz and $\omega T_1=\pi $ (blue, circles), $2\pi $ (red, crosses) and $3\pi$(green, squares). Each marker corresponds to a measurement.} \label{fig:collapsed}
\end{figure}
The final question which remains to be answered is whether the exponent $\bar\chi$ depends on the frequency of the measurements or not. For that we have fixed the coupling strength and explored three values of the period, $T_1,$ studying the average exponential behavior. The result is shown in Fig.~\ref{fig:collapsed}, collapsing all numerical simulations in the dimensionless quantity $t/T_1,$ and finding that they have very similar slopes.

\subsection{Weak measurements}

So far we have considered ideal projective measurements, introducing only some stochasticity in the time at which the measurement event is produced. We will now add another ingredient to our measurement model, which is the possibility that the detector only performs a partial measurement, leaving the state ``untouched'' with a nonzero probability, $\epsilon.$

We can easily model an imperfect detector using the formalism of completely positive maps, operations that transform density matrices into density matrices. If $\rho$ and $\rho'$ are the states of the qubit-resonator system before and after the measurement, we will write, up to normalization
\begin{equation}
\rho'= (1-\epsilon) (\openone - \hat{P}_e) \rho  (\openone - \hat{P}_e) + \epsilon \rho. \label{eq:errors}
\end{equation}
This is read as follows. With probability $\epsilon$ the measurement device will do nothing, leaving the state untouched. With probability $(1-\epsilon)$ the measurement device will detect the state of the qubit. In this case it will either give us a positive signal, moment at which we will stop the experiment, or it will not produce anything at all, and we will continue with the projected state $(\openone-\hat{P}_e)\rho(\openone-\hat{P}_e),$ that has the qubit deexcited, $\ket{g}.$

This qualitative model describes measurements from a SQUID \cite{measurementsqubits1,measurementqubits2}, where we place ourselves on the verge of metastability and assume that if the qubit is in the excited state, $\ket{e},$ the SQUID will tunnel to the voltage state with probability $(1-\epsilon),$ giving no signal for $\ket{g}.$ Note that with probability $\epsilon$ the SQUID may not tunnel and then we will gain no information about the qubit or the resonator.

\begin{figure}[t]
  \centering
  \includegraphics[width=0.95\linewidth]{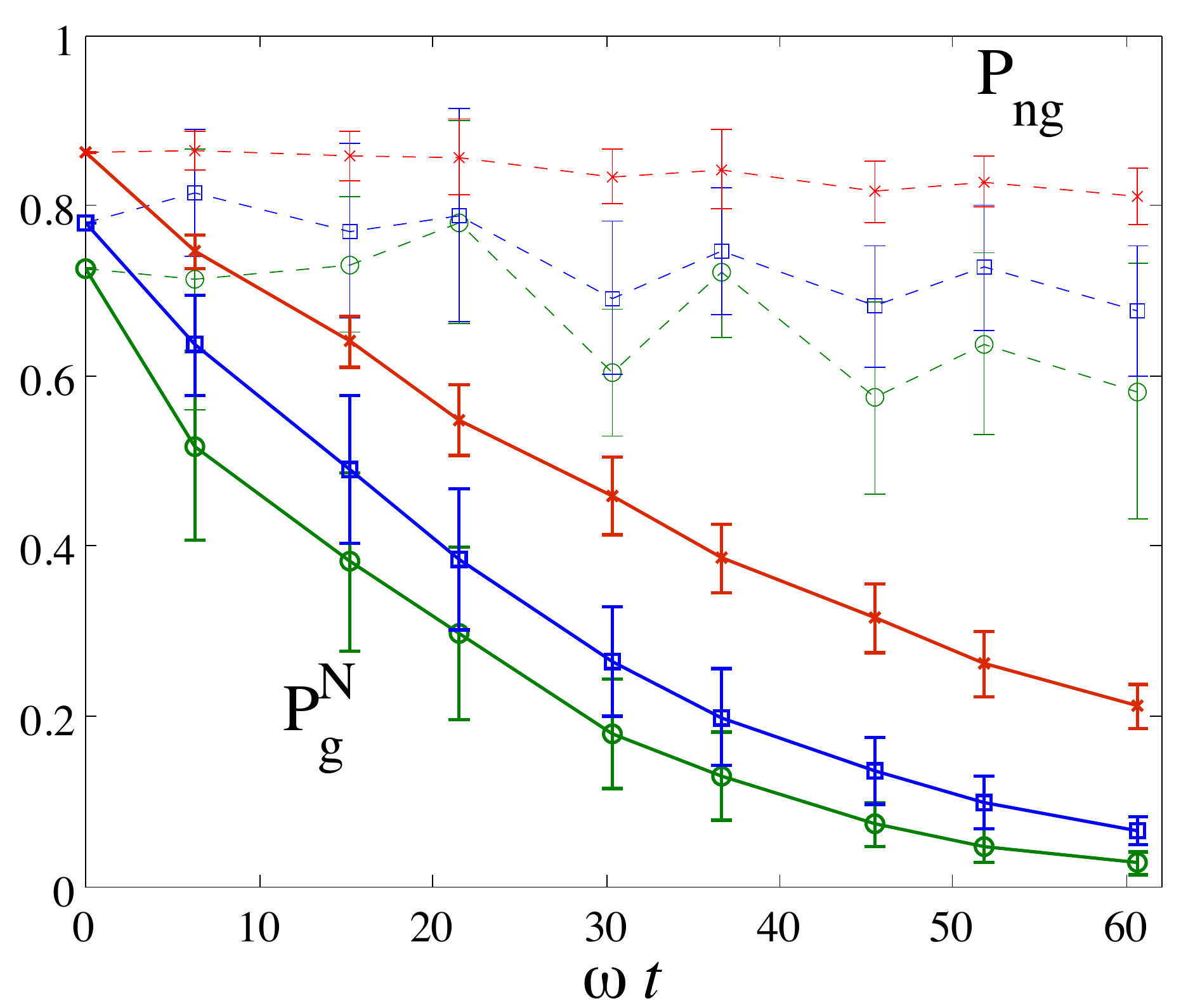}
  \caption{(Color online) Survival probability $P_g^N$ (solid lines) and $p_{ng}$ (dashed lines) vs. $\omega t,$ with $g=\omega=\omega_0= 1$ GHz , $\omega T_1=2\pi$ and $\epsilon=0.2, 0.1$ and $0$ (crosses, squares and circles, respectively). The probability of a measurement at a time $\omega t=\omega t_0$ is averaged over 20 random values  within the interval $[\omega t_0-0.2\pi,\omega t_0+0.2\pi]$} \label{fig:8Mtimeepsilon} 
\end{figure}

In Fig. \ref{fig:8Mtimeepsilon} we analyze the impact of $\epsilon$ in our previous results. Even for large errors $\epsilon=0.2$ we retain the exponential behavior observed in Fig. \ref{fig:8Mtime}a, with acceptable error bars that decrease with increasing number of measurements --- in other words, the qubit is still efficiently projected to the excited state.
  
\subsection{Relaxation and dephasing}

Throughout this work we have considered in the numerical simulations the model given by the Hamiltonian in Eq. (\ref{eq:hamiltonian}) which do not include effects like relaxation or qubit dephasing, usually included in master equation approaches.

We want to remark that it is still an open question, both experimentally and theoretically, to understand and model the dissipation and decoherence processes of quantum circuits in the presence of ultrastrong qubit-cavity coupling. One popular approach~\cite{werlang08,dodonov10} is to combine the usual photon leakage mechanism from quantum optics models, $\mathcal{L}(\rho)\sim 2 a\rho a^\dagger - a^\dagger a \rho - \rho a^\dagger a,$ with the qubit-cavity Hamiltonian. Note that in such a combination, the asymptotic states of the dissipation (the vacuum) and of the interaction (populated cavity) are incompatible, and one may find excitations induced by the dissipative terms, an infinite stream of photons leaking out of the cavity and other controversial phenomena.

These effects disappear when one rederives the master equation from first principles, using the qubit-cavity eigenstates of the ultrastrong coupling model and the usual zero temperature baths. In the resulting models the main relaxation mechanisms are found to be the decay to the ground state $\ket{G}$ and a dephasing of the joint cavity-qubit states ---in other words, dissipation and decoherence in the proper basis---. If we assume this reasonable model, then we can conclude that the exponential laws derived in this manuscript are not significantly distorted. To begin with, relaxation to the ground state $\ket{G}$ just makes the experiment closer to the truncated Hilbert space model considered in Sect.~\ref{sec:repeated}, and in particular to the exponential law from Eq.~(\ref{eq:truncated-exp}). For strong couplings, decoherence amounts to random modulations of the qubit-cavity energy levels, without significantly affecting the populations, $|c_i|^2.$ Since this is the most relevant quantity in all the previous discussions, we can also expect that, up to minor changes in the rates, the anti-Zeno effect will also survive.

\section{Conclusions}

We have considered a system consisting in a superconducting qubit coupled to a closed transmission line, operating in the ultrastrong coupling regime. The ground state in such scheme is not just a product of the ground states of the qubit and the cavity, as is the case for weaker couplings. On the contrary, the vacuum of the system is dressed by the interaction and so it contains a relevant probability of finding the qubit excited. This probability is proportional to the square of the coupling strength. We have introduced a protocol for detecting that excitations with certainty, maximizing the small probabilities that are obtained with only one measurement.

Our main result is that, after a number of periodic measurements of the qubit, the probability of finding it in the ground state in all the measurements goes exponentially to zero, even if the measurements are weak and are performed with a slow repetition rate in comparison with the fast dynamics of the interaction. We refer to this as slow quantum anti-Zeno effect. Like the well known quantum anti-Zeno effect, the result is the acceleration of a transition, in this case the exotic transition $\ket g\rightarrow \ket e,$ which becomes relevant in this regime due to the breakdown of the RWA. But this procedure is less experimentally demanding, since it requires a smaller number of measurements and a shorter duration of the period at which they are performed. We have shown that the protocol is robust to large errors in the measurement process, when a realistic SQUID readout is considered.

This is one of the first experimentally accessible consequences of the new ultrastrong coupling regime and can only be derived beyond the RWA. The physical nature of the ground state qubit self-excitations, commonly considered as  a virtual process without possible experimental record, seems now to be clear. Moreover, although the ultrastrong coupling entails a very fast dynamics, we have shown that valuable information of the interaction can be extracted efficiently with the current slow and imperfect measurement technologies.

Finally, we want to remark that strong qubit excitations have also been found theoretically in models that combine the full Rabi coupling with traditional dissipative contributions~\cite{werlang08,dodonov10}. However, the form of those dissipative terms is questionable in non-RWA setups, and furthermore, there is no justification to equate the sparse measurement setup in this work to a particular dissipative model. This lack of equivalence between models manifests in the fact that, as we have seen numerically, the sparsely repeated measurements can hit certain resonances that invalidate the anti-Zeno dynamics.
 
 \acknowledgements

The authors would like to thank Enrique Solano and Daniel Ballester for useful discussions. This work is supported by Spanish MICINN Projects FIS2008-05705 and FIS2009-10061, and CAM research consortium QUITEMAD S2009-ESP-1594.

\end{document}